\documentclass[aps,preprint,nofootinbib,superscriptaddress]{revtex4}%
\usepackage{amssymb}
\usepackage{amsmath}
\usepackage{epsfig}
\usepackage{amsfonts}
\usepackage{graphicx}
\usepackage{hyperref}%
\providecommand{\U}[1]{\protect\rule{.1in}{.1in}}
\begin{document}
\title{Biquaternions and ADHM Construction of Non-Compact SL(2,C) Yang-Mills Instantons}
\author{}
\author{Sheng-Hong Lai}
\email{xgcj944137@gmail.com}
\affiliation{Department of Electrophysics, National Chiao-Tung University and Physics
Division, National Center for Theoretical Sciences, Hsinchu, Taiwan, R.O.C.}
\author{Jen-Chi Lee}
\email{jcclee@cc.nctu.edu.tw}
\affiliation{Department of Electrophysics, National Chiao-Tung University and Physics
Division, National Center for Theoretical Sciences, Hsinchu, Taiwan, R.O.C.}
\author{I-Hsun Tsai}
\email{ihtsai@math.ntu.edu.tw}
\affiliation{Department of Mathematics, National Taiwan University, Taipei, Taiwan, R.O.C. }
\author{}
\date{\today }

\begin{abstract}
We extend quaternion calculation in the ADHM construction of $Sp(1)$
($=SU(2)$) self-dual Yang-Mills (SDYM) instantons to the case of biquaternion.
We use the biconjugate operation of biquaternion first introduced by Hamilton
to construct the non-compact $SL(2,C)$ $k$-instantons. The number of moduli
for $SL(2,C)$ $k$-instantons is found to be twice of that of $Sp(1)$, $16k-6$.
These new $SL(2,C)$ instanton solutions contain the $SL(2,C)$ $(M,N)$
instanton solutions constructed previously as a subset. The structures of
singularities or jumping lines of the complete $SL(2,C)$ $k=1,2,3$ instantons
with $10,26,42$ moduli parameters are particularly investigated. The existence
of singular structures of the $SL(2,C)$ $k$-instantons is mathematically
consistent with recent results of solutions of complex ADHM equations. It may
also help to clearify the long standing global singularity problems associated
with Backlund transformations of $SU(2)$ instantons.

\end{abstract}
\preprint{ }
\maketitle
\tableofcontents
%

\setcounter{equation}{0}
\renewcommand{\theequation}{\arabic{section}.\arabic{equation}}%

\section{\bigskip Introduction}

The discovery of classical exact solutions of Euclidean $SU(2)$
(anti)self-dual Yang-Mills (SDYM) equation was one of the most important
achievements in the developements of both quantum field theory and algebraic
geometry in 1970's. The first BPST $1$-instanton solution \cite{BPST} with $5$
moduli parameters was found in 1975. Soon later the CFTW k-instanton solutions
\cite{CFTW} with $5k$ moduli parameters were constructed, and then the number
of moduli parameters of the solutions for each homotopy class $k$ was extended
to $5k+4$ ($5$,$13$ for $k=1$,$2$) \cite{JR} based on the consideration of
$4D$ conformal symmetry of massless pure YM equation. The complete solutions
with $8k-3$ moduli parameters for each $k$-th homotopy class were finally
worked out in 1978 by mathematicians ADHM \cite{ADHM} using method in
algebraic geometry. By using an one to one correspondence between
anti-self-dual SU(2)-connections on $S^{4}$ and certain holomorphic vector
bundles of rank two on $CP^{3}$, ADHM converted the highly nontrivial system
of non-linear partial differential equations of anti-SDYM into a much more
simpler system of quadratic algebraic equations in quaternions. The explicit
closed form of the complete solutions for $k=2,3$ had been worked out
\cite{CSW}.

Many interesting further developments, including supersymmetric YM instantons
\cite{28}, Heterotic string instantons \cite{Stromin} and noncommutative YM
instantons \cite{Harvey}etc., followed since then. One important application
of instantons in algebraic geometry was the classification of four-manifolds
\cite{5}. On the physics side, the non-perturbative instanton effect in QCD
resolved the long standing $U(1)_{A}$ problem \cite{U(1)}. On the other hand,
another important application of YM instantons in quantum field theory was the
introduction of $\theta$- vacua \cite{the} in nonperturbative QCD, which
created the strong $CP$ problem. This unsolved issue remains a puzzle till
even today.\bigskip

In addition to $SU(2)$, the ADHM construction has been generalized to the
cases of $SU(N)$ SDYM and many other SDYM theories with compact Lie groups
\cite{CSW,JR2}. In this paper we are going to consider the classical solutions
of non-compact $SL(2,C)$ SDYM system. $SL(2,C)$ YM theory was first discussed
by some authors in 1970's \cite{WY,Hsu}. They found out that the complex
$SU(2)$ YM field configurations can be interpreted as the real field
configurations in $SL(2,C)$ YM theory. However, due to the non-compactness of
$SL(2,C)$, the Cartan-Killing form or group metric of $SL(2,C)$ is not
positive definite. Thus the action integral and the Hamiltonian of non-compact
$SL(2,C)$ YM theory may not be positve. Nevertheless, there are still
important motivations to study $SL(2,C)$ SDYM theory. It was shown that the
$4D$ $SL(2,C)$ SDYM equation can be dimensionally reduced to many important
$1+1$ dimensional integrable systems \cite{Mason}, such as the KdV equation
and the nonlinear Schrodinger equation. In 1985 \cite{Ward}, it was even
conjectured by Ward that many (and perhaps all?) integrable or solvable
equations may be obtained from the SDYM equations (or its generalizations) by reduction.

On the other hand, the parametric Backlund transformation (PBT) constructed in
terms of $J$-matrix formulation \cite{J} of $SU(2)$ Yang-Mills theory takes a
real $SU(2)$ gauge field into the real $SU(1,1)$ gauge field and vice versa
\cite{PBT,AW}. Therefore it would be of interest to study $SL(2,C)$ gauge
group which contains the non-compact subgroup $SU(1,1)$ as well as the compact
subgroup $SU(2)$, and the solutions to the $SL(2,C)$ SDYM can be transformed
into the new ones by any arbitrary numbers of PBT. Moreover, as it will turn
out, there are singularities which can not be gauged away in the field
configurations of $SL(2,C)$ YM instantons. This may help to clearify the long
standing issue of global singularity problems associated with Backlund
transformations \cite{PBT,AW} of $SU(2)$ SDYM instantons. More recently the
$SL(2,C)$ SDYM theory including its singular structure was also considered in
the literatures from mathematical point of view \cite{math1,math3,math2}.

In 1984 \cite{Lee}, some exact solutions of $SL(2,C)$ SDYM system were
explicitly constructed in the $(R,\bar{R})$-gauge, which was a direct
generalization of $R$-gauge in Yang's formulation \cite{Yang} of $SU(2)$ SDYM
equation. The topological charges of these so-called $(M,N)$ solutions
\cite{Lee} were calculated by the third homotopy group $\pi_{3}(SL(2,C)=Z$. In
this paper, we extend quaternion calculation in the ADHM construction of
compact $Sp(1)$ (and $SU(N)$, $Sp(N)$, $O(N)$ cases) SDYM instantons to the
case of biquaternion of Hamilton \cite{Ham}. We will use the biconjugate
operation of biquaternion first introduced by Hamilton \cite{Ham} to construct
the $SL(2,C)$ SDYM instantons. These new $SL(2,C)$ instanton solutions contain
previous $SL(2,C)$ $(M,N)$ instanton solutions as a subset constructed in
1984. In addition, we will obtain many more new $SL(2,C)$ SDYM field
configurations. It turns out that the number of moduli for solutions of the
$SL(2,C)$ SDYM for each $k$-th homotopy class is twice of that of the case of
$SU(2)$ SDYM, namely $16k-6$.

This paper is organized by the following. In section II, we set up the
formalism of $SL(2,C)$ SDYM theory and review the previous $(M,N)$ instanton
solutions \cite{Lee}. Section III is devoted to the general construction of
solutions with $16k-6$ parameters by using biquaternions. Three explicit
examples will be given in section IV. These include the $(M,N)$ instanton
solutions, the complete $k=2,3$ instanton solutions and a detailed discussion
of $1$-instanton solution and the structure of its singularities depending on
its moduli space with$10$ parameters. The singular structures of $SL(2,C)$
$k$-instantons will also be discussed in section IV. The singularities called
"jumping \ lines" of $1$-instantons are intersections of zeros of $P_{2}(x)$
and $P_{1}(x)$ polynomials of $4$ variables with degrees $2$ and $1$
respectively. For singularities of general $k$-instanton field configurations,
one encounters intersections of zeros of $P_{2k}(x)$ and $P_{2k-1}(x)$
polynomials with degrees $2k$ and $2k-1$ respectively. In particular, the
complete jumping lines of $SL(2,C)$ $k=1,2,3$ instantons with $10,26,42$
moduli parameters are calculated. The existence of singular structure of the
non-compact $SL(2,C)$ SDYM field configurations discovered in this paper is
consistent with the recent use of "sheaves" by Frenkel-Jardim \cite{math2} for
complex ADHM equations, rather than just the restricted notion of "vector
bundles". Finally, a brief conclusion is given in section V.

\section{\bigskip Review of $SL(2,C)$ $(M,N)$ Instantons}

In this section, we will use the convention $\mu=1,2,3,4$ and $\epsilon
_{1234}=1$ for $4D$ Euclidean space. We will first briefly review the
$SL(2,C)$ solutions constructed 30 years ago in \cite{Lee}. Wu and Yang
\cite{WY} have shown that there are two linearly independent choices of
$SL(2,C)$ group metric\newline%

\begin{equation}
g^{a}=%
\begin{pmatrix}
I & 0\\
0 & -I
\end{pmatrix}
,g^{b}=%
\begin{pmatrix}
0 & I\\
I & 0
\end{pmatrix}
\end{equation}
where $I$ is the $3\times3$ unit matrix. In general, we can choose
\begin{equation}
g=\cos\theta g^{a}+\sin\theta g^{b}%
\end{equation}
where $\theta$ = real constant. Note that the metric is not positive definite
due to the non-compactness of $SL(2,C).$ On the other hand, it was shown that
$SL(2,C)$ group can be decomposed such that \cite{Lee}%
\begin{equation}
SL(2,C)=SU(2)\cdot P,P\in H\newline%
\end{equation}
\newline where $SU(2)$ is the maximal compact subgroup of $SL(2,C)$, $P\in H$
(not a group) and $H=\{P|P$ is Hermitain, positive definite, and $detP=1\}$.
The parameter space of $H$ is a noncompact space $R^{3}$. The third homotopy
group is thus \cite{Lee}%
\begin{equation}
\pi_{3}[SL(2,C)]=\pi_{3}[S^{3}\times R^{3}]=\pi_{3}(S^{3})\cdot\pi_{3}%
(R^{3})=Z\cdot I=Z\newline\newline%
\end{equation}
\newline where $I$ is the identity group, and $Z$ is the integer
group.\newline\qquad Wu and Yang \cite{WY} have shown that a complex $SU(2)$
gauge field is related to a real $SL(2,C)$ gauge field. Starting from $SU(2)$
complex gauge field formalism, we can write down all the $SL(2,C)$ field
equations. Let
\begin{equation}
G_{\mu}^{a}=A_{\mu}^{a}+iB_{\mu}^{a}%
\end{equation}
and, for convenience, we set the coupling constant $g=1$. The complex field
strength is defined as
\begin{equation}
F_{\mu\nu}^{a}\equiv H_{\mu\nu}^{a}+iM_{\mu\nu}^{a},a,b,c=1,2,3
\end{equation}
where
\begin{align}
H_{\mu\nu}^{a}  &  =\partial_{\mu}A_{\nu}^{a}-\partial_{\nu}A_{\mu}%
^{a}+\epsilon^{abc}(A_{\mu}^{b}A_{\nu}^{c}-B_{\mu}^{b}B_{\nu}^{c}),\nonumber\\
M_{\mu\nu}^{a}  &  =\partial_{\mu}B_{\nu}^{a}-\partial_{\nu}B_{\mu}%
^{a}+\epsilon^{abc}(A_{\mu}^{b}B_{\nu}^{c}-A_{\mu}^{b}B_{\nu}^{c}),
\end{align}
then Yang-Mills equation can be written as
\begin{align}
\partial_{\mu}H_{\mu\nu}^{a}+\epsilon^{abc}(A_{\mu}^{b}H_{\mu\nu}^{c}-B_{\mu
}^{b}M_{\mu\nu}^{c})  &  =0,\nonumber\\
\partial_{\mu}M_{\mu\nu}^{a}+\epsilon^{abc}(A_{\mu}^{b}M_{\mu\nu}^{c}-B_{\mu
}^{b}H_{\mu\nu}^{c})  &  =0.
\end{align}
The $SL(2,C)$ SDYM equations are%
\begin{align}
H_{\mu\nu}^{a}  &  =\frac{1}{2}\epsilon_{\mu\nu\alpha\beta}H_{\alpha\beta
},\nonumber\\
M_{\mu\nu}^{a}  &  =\frac{1}{2}\epsilon_{\mu\nu\alpha\beta}M_{\alpha\beta}.
\label{self}%
\end{align}
Yang-Mills Equation can be derived from the following Lagrangian%
\begin{equation}
L_{\theta}=\frac{1}{4}[F_{\mu\nu}^{i}]^{T}g_{ij}[F_{\mu\nu}^{j}]=\cos{\theta
}(\frac{1}{4}H_{\mu\nu}^{a}H_{\mu\nu}^{a}-\frac{1}{4}M_{\mu\nu}^{a}M_{\mu\nu
}^{a})+\sin{\theta}(\frac{1}{2}H_{\mu\nu}^{a}M_{\mu\nu}^{a}) \label{action}%
\end{equation}
where $F_{\mu\nu}^{k}=H_{\mu\nu}^{k}$ and $F_{\mu\nu}^{3+k}=M_{\mu\nu}^{k}$
for $k=1,2,3$. Note that $L_{\theta}$ is indefinite for any real value
$\theta$. We shall only consider the particular case for $\theta=0$ in this
section, i.e.
\begin{equation}
L=\frac{1}{4}(H_{\mu\nu}^{a}H_{\mu\nu}^{a}-M_{\mu\nu}^{a}M_{\mu\nu}^{a}),
\end{equation}
for the action density in discussing the homotopic classifications of our solutions.

In the Yang formulation of $SU(2)$ SDYM theory, one first performs analytic
continuation of $x_{\mu}$ to complex space, the self-dual condition
Eq.(\ref{self}) is still valid in complex space. We then perform the following
transformations in complex space \cite{Yang}
\begin{align}
\sqrt{2}y=x_{1}+ix_{2},  &  \sqrt{2}\bar{y}=x_{1}-ix_{2},\nonumber\\
\sqrt{2}z=x_{3}-ix_{4},  &  \sqrt{2}\bar{z}=x_{3}+ix_{4},\\
\sqrt{2}G_{y}=G_{1}-iG_{2},  &  \sqrt{2}G_{\bar{y}}=G_{1}+iG_{2},\nonumber\\
\sqrt{2}G_{z}=G_{3}+iG_{4},  &  \sqrt{2}G_{\bar{z}}=G_{3}-iG_{4}.
\end{align}
Note that $y$ and $\bar{y}$ (similarly $z$ and $\bar{z}$) are independent
complex numbers. They are complex conjugate to each other when we restrict
$x_{\mu}$ to be real. The self-dual equation then reduces to
\begin{align}
F_{yz}=F_{\bar{y}\bar{z}}=0,  & \label{dual}\\
F_{y\bar{y}}+F_{z\bar{z}}=0.  &  \label{J}%
\end{align}
Eq.(\ref{dual}) is now in the pure gauge and can be integrated once. In the
so-called R-gauge, Eq.(\ref{J}) reduces to \cite{Yang}%
\begin{align}
\phi\lbrack\phi_{y\bar{y}+\phi_{z\bar{z}}}]-\phi_{y}\phi_{\bar{y}}-\phi
_{z}\phi_{\bar{z}}+\rho_{y}\bar{\rho}_{\bar{y}}+\rho_{z}\bar{\rho}_{\bar{z}}
&  =0,\nonumber\\
\phi\lbrack\rho_{y\bar{y}}+\rho_{z\bar{z}}]-2\rho_{y}\phi_{\bar{y}}-2\rho
_{z}\phi_{\bar{z}}  &  =0,\nonumber\\
\phi\lbrack\bar{\rho}_{y\bar{y}}+\bar{\rho}_{z\bar{z}}]-2\bar{\rho}_{\bar{y}%
}\phi_{y}-2\bar{\rho}_{\bar{z}}\phi_{z}  &  =0, \label{R}%
\end{align}
where $\phi$, $\rho$ and $\bar{\rho}$ are three independent complex valued
functions or six real valued functions. For the case of $SU(2)$, one needs to
impose the reality conditions $\phi\doteq$real, $\bar{\rho}\doteq\rho^{\ast}$
so that $G_{\mu}$ will be a real gauge field. Here $"\doteq"$ means $"="$ when
we restrict $x_{\mu}$ to be real. For the case of $SL(2,C)$ considered in this
paper, we drop out the reality conditions and the $R$-gauge will be called
$(R,\bar{R})$ gauge. Thus in the $SL(2,C)$ $(R,\bar{R})$ gauge, $G_{\mu}$ can
be complex and there are three independent complex valued functions or six
real valued functions. It is easily seen that one set of solutions of
Eq.(\ref{R}) is
\begin{equation}
\rho_{y}=\phi_{\bar{z}},\,\rho_{z}=-\phi_{\bar{y}},\,\bar{\rho}_{\bar{y}%
}=\,\phi_{z},\bar{\rho}_{\bar{z}}=-\phi_{y}.
\end{equation}
For the $SL(2,C)$ case, this is to say that the complex gauge potential
$G_{\mu\nu}^{a}$ can be taken as
\begin{equation}
G_{\mu\nu}^{a}=-\bar{\eta}_{\mu\nu}^{a}\partial_{\nu}(\ln{\phi}) \label{CFTW}%
\end{equation}
where $\bar{\eta}_{\mu\nu}^{a}$ is defined to be \cite{U(1)}
\begin{subequations}
\begin{align}
\eta_{\mu\nu}^{a}  &  =\eta^{a\mu\nu}=\epsilon^{a\mu\nu4}+\delta^{a\mu}%
\delta^{\nu4}-\delta^{a\nu}\delta^{\mu4},\\
\bar{\eta}_{\mu\nu}^{a}  &  =\bar{\eta}^{a\mu\nu}=(-1)^{(\delta_{\mu_{4}%
}+\delta_{\nu_{4}})}\eta^{a\mu\nu}.
\end{align}
Eq.(\ref{CFTW}) is the Corrigan-Fairlie-'t Hooft-Wilczek (CFTW) \cite{CFTW}
anastz which is used to obtain $SU(2)$ $k$-instanton solutions. But for the
case of $SL(2,C)$, $\phi$ is a complex-valued function. Substitution of
Eq.(\ref{CFTW}) into Eq.(\ref{self}) and using \cite{U(1)}%

\end{subequations}
\begin{subequations}
\begin{align}
\eta_{a\mu\nu}=\frac{1}{2}\epsilon_{\mu\nu\alpha\beta}\eta_{a\alpha\beta}%
,\bar{\eta}_{a\mu\nu}=-\frac{1}{2}\epsilon_{\mu\nu\alpha\beta}\bar{\eta
}_{a\alpha\beta},  & \label{3-4a}\\
\delta_{\kappa\lambda}\eta_{a\mu\nu}+\delta_{\kappa\nu}\eta_{a\lambda\nu
}+\delta_{\kappa\mu}\eta_{a\nu\lambda}+\eta_{a\sigma\kappa}\epsilon
_{\lambda\mu\nu\sigma}=0,  & \\
\epsilon_{abc}\eta_{b\mu\nu}\eta_{c\kappa\lambda}=\delta_{\mu\kappa}\eta
_{a\nu\lambda}-\delta_{\mu\lambda}\eta_{a\nu\kappa}-\delta_{\nu\kappa}%
\eta_{a\mu\lambda},  &
\end{align}
we obtain
\end{subequations}
\begin{subequations}
\begin{equation}
\frac{1}{\phi}\Box\phi=0 \label{3-5}%
\end{equation}
where $\Box=\partial_{\mu}\partial_{\mu}=2(\partial_{y}\partial_{\bar{y}%
}+\partial_{z}\partial_{\bar{z}}).$ Note that for $\phi=p+iq,$
\end{subequations}
\begin{equation}
\frac{1}{p}\Box p=0,\;\frac{1}{q}\Box q=0 \label{3-6}%
\end{equation}
satisfy Eq.(\ref{3-5}). Eq.(\ref{3-6}) has the following solutions \cite{Lee}
\begin{align}
p  &  =1+\sum_{i=1}^{M}\frac{\alpha_{i}^{2}}{|x_{\mu}-a_{i\mu}|^{2}%
},0,\nonumber\\
q  &  =1+\sum_{i=1}^{M}\frac{\beta_{i}^{2}}{|x_{\mu}-b_{j\mu}|^{2}},0
\end{align}
where $\alpha_{i},\beta_{j}$ are real constants, $a_{i\mu},b_{j\mu}$ are real
constant $4$-vector. A special case is that when $p=q$ $(M=N,\alpha_{i}%
=\beta_{j},a_{i\mu}=b_{j\mu})$ or $q=0,p\neq0$ or $p=0,q\neq0$, the $SU(2)$
CFTW $k$-instanton solutions can be embedded in that of $SL(2,C)$ gauge field.
In general, we have the pure $SL(2,C)$ solutions
\begin{equation}
G_{\mu}^{a}=-\bar{\eta}_{\mu\nu}^{a}\partial(\ln{\phi})=-\bar{\eta}_{\mu\nu
}^{a}\frac{1}{p^{2}+q^{2}}[pp_{\nu}+qq_{\nu}+i(pq_{\nu}-qp_{\nu})].
\end{equation}

For the simplest $SL(2,C)$ $1$-instanton case $(M,N)=(1,0)$, let's take
\begin{equation}
M=1,N=0,p=1+\frac{\alpha_{1}^{2}}{y^{2}},q=1 \label{case1}%
\end{equation}
where $y_{\mu}\equiv\,x_{\mu}-a_{1\mu},y^{2}\equiv y_{\mu}y_{\mu}$, the gauge
potentials can be calculated to be
\begin{align}
A_{\mu}^{a}  &  =\bar{\eta}_{\mu\nu}^{a}y_{\nu}\frac{2\alpha_{1}^{2}%
(y^{2}+\alpha_{1}^{2})}{y^{2}[y^{4}+(y^{2}+\alpha_{1}^{2})^{2}]},\nonumber\\
B_{\mu}^{a}  &  =-\bar{\eta}_{\mu\nu}^{a}y_{\nu}\frac{2\alpha_{1}^{2}}%
{y^{4}+(y^{2}+\alpha_{1}^{2})^{2}}. \label{(1,0)}%
\end{align}
The gauge potential $A_{\mu}^{a}$ has a singularity at $x_{\mu}=a_{1\mu}$
which is a gauge artifact that can be gauged away by a $SL(2,C)$ gauge
transformation. Define
\begin{equation}
U_{1}(x)=\frac{(x_{4}+ix_{j}\sigma_{j})}{|x|}=\hat{x}_{\mu}S_{\mu},U_{1}(x)\in
SU(2)\subset SL(2,C) \label{SU}%
\end{equation}
where $S_{1,2,3}=i\sigma_{1,2,3}$. After making a large gauge transformation
by $U_{1}(x)$, we have \cite{Lee}
\begin{align}
A_{\mu}^{^{\prime}a}  &  =\eta_{\mu\nu}^{a}y_{\nu}\frac{2(2y^{2}+\alpha
_{1}^{2})}{y^{4}+(y^{2}+\alpha_{1}^{2})^{2}},\nonumber\\
B_{\mu}^{\prime a}  &  =\eta_{\mu\nu}^{a}y_{\nu}\frac{2\alpha_{1}^{2}}%
{y^{4}+(y^{2}+\alpha_{1}^{2})^{2}}, \label{1,0}%
\end{align}
which are regular $SL(2,C)$ solution. The corresponding field strength can be
calculated to be \cite{Lee}
\begin{align}
H_{\mu\nu}^{\prime a}  &  =-4\eta_{\mu\nu}^{a}\alpha_{1}^{2}\frac
{2y^{4}+4\alpha_{1}^{2}y^{2}+\alpha_{1}^{4}}{[y^{4}+(y^{2}+\alpha_{1}^{2}%
)^{2}]^{2}},\nonumber\\
M_{\mu\nu}^{\prime a}  &  =-4\eta_{\mu\nu}^{a}\alpha_{1}^{2}\frac
{2y^{4}-\alpha_{1}^{4}}{[y^{4}+(y^{2}+\alpha_{1}^{2})^{2}]^{2}}, \label{1,0f}%
\end{align}
which are self-dual by Eq.(\ref{3-4a}).

Alternatively, instead of taking Eq.(\ref{case1}), let's take $(M,N)=(0,1)$
\begin{equation}
M=0,\,N=1,\,p=1,\,q=1+\frac{\beta_{1}^{2}}{y^{2}}, \label{case2}%
\end{equation}
where$\;y_{\mu}\equiv x_{\mu}-b_{1\mu},\,y^{2}\equiv y_{\mu}y_{\mu}.$ Then we
have%
\begin{equation}
\phi=1+i+\frac{i\beta_{1}^{2}}{|x-y_{1}|^{2}}. \label{(0,1)}%
\end{equation}
It can be shown that for $SU(2)$ complex YM equation with a complex source
term $J_{\mu}$, the complex gauge potential for $(M,N)$ solution is related to
the complex conjugate of $(N,M)$ solution with $J_{\mu}$ replaced by $J_{\mu
}^{\ast}$. For the present pure YM case without $J_{\mu}$, it can be shown
that Eq.(\ref{case2}) leads to a solution which is equivalent to the solution
in Eq.(\ref{(1,0)}). We will see this equivalence in section IV where more
general $1$-instanton solution will be constructed. In general, one can
generalize the $1$-instanton solution to the k-instanton cases. For the
multi-instanton solutions, say $k=2$ for example, we get%
\begin{equation}
\phi=(1+i+\frac{\alpha_{1}^{2}}{|x-y_{1}|^{2}}+\frac{i\beta_{1}^{2}}%
{|x-y_{2}|^{2}}). \label{(1,1)}%
\end{equation}
In general, the topological charge of the $(M,N)$ solution was found to be
$Q=M+N$ \cite{Lee}. For the boundary condistions%
\begin{equation}
\lim_{r\rightarrow\infty}H_{\mu\nu}^{a}=\lim_{r\rightarrow\infty}M_{\mu\nu
}^{a}=0,
\end{equation}
the action integral for the case of $\theta=0$ in Eq.(\ref{action}) can be
calculated to be \cite{Lee}%
\begin{align}
\int_{\mathbb{R}^{4}}\mathrm{d}^{4}xL  &  =\int_{\mathbb{R}^{4}}\mathrm{d}%
^{4}x\frac{1}{4}(H_{\mu\nu}^{a}H_{\mu\nu}^{a}-M_{\mu\nu}^{a}M_{\mu\nu}%
^{a})\nonumber\\
&  =8\pi^{2}Q\,=\,8\pi^{2}(M+N).
\end{align}
Note that for the non-compact $SL(2,C)$ case, unlike the $SU(2)$ case, there
is no proof that instanton action is the minimum action in each homotopy class.

\section{Biquaternions and $SL(2,C)$ ADHM YM Instantons}

\bigskip In this section and section IV, in contrast to the last section, we
will use the convention $\mu=0,1,2,3$ and $\epsilon_{0123}=1$ for $4D$
Euclidean space. Instead of quaternion in the $Sp(1)$ ($=SU(2)$) ADHM
construction, we will use \textit{biquaternion} to construct $SL(2,C)$ SDYM
instantons. A quaternion $x$ can be written as%
\begin{equation}
x=x_{\mu}e_{\mu}\text{, \ }x_{\mu}\in R\text{, \ }e_{0}=1,e_{1}=i,e_{2}%
=j,e_{3}=k \label{x}%
\end{equation}
where $e_{1},e_{2}$ and $e_{3}$ anticommute and obey%
\begin{align}
e_{i}\cdot e_{j}  &  =-e_{j}\cdot e_{i}=\epsilon_{ijk}e_{k};\text{
\ }i,j,k=1,2,3,\\
e_{1}^{2}  &  =-1,e_{2}^{2}=-1,e_{3}^{2}=-1.
\end{align}
The conjugate quarternion is defined to be%

\begin{equation}
x^{\dagger}=x_{0}e_{0}-x_{1}e_{1}-x_{2}e_{2}-x_{3}e_{3}%
\end{equation}
so that the norm square of a quarternion is%
\begin{equation}
|x|^{2}=x^{\dagger}x=x_{0}^{2}+x_{1}^{2}+x_{2}^{2}+x_{3}^{2}. \label{norm}%
\end{equation}
Occasionaly the unit quarternions were expressed as Pauli matrices%
\begin{equation}
e_{0}\rightarrow%
\begin{pmatrix}
1 & 0\\
0 & 1
\end{pmatrix}
,e_{i}\rightarrow-i\sigma_{i}\ \text{; }i=1,2,3.
\end{equation}

A (ordinary) biquaternion (or complex-quaternion) $z$ can be written as%
\begin{equation}
z=z_{\mu}e_{\mu}\text{, \ }z_{\mu}\in C,
\end{equation}
which will be used in this paper. Occasionally $z$ can be written as%
\begin{equation}
z=x+yi
\end{equation}
where $x$ and $y$ are quaternions and $i=\sqrt{-1},$ not to be confused with
$e_{1}$ in Eq.(\ref{x}). There are two other types of biquaternions in the
literature, the split-biquaternion and the dual biquaternion. For
biquaternion, Hamilton introduced two types of conjugations, the biconjugation
\cite{Ham}%
\begin{equation}
z^{\circledast}=z_{\mu}e_{\mu}^{\dagger}=z_{0}e_{0}-z_{1}e_{1}-z_{2}%
e_{2}-z_{3}e_{3}=x^{\dagger}+y^{\dagger}i,
\end{equation}
which will be heavily used in this paper, and the complex conjugation%
\begin{equation}
z^{\ast}=z_{\mu}^{\ast}e_{\mu}=z_{0}^{\ast}e_{0}+z_{1}^{\ast}e_{1}+z_{2}%
^{\ast}e_{2}+z_{3}^{\ast}e_{3}=x-yi.
\end{equation}
In contrast to Eq.(\ref{norm}), the norm square of a biquarternion used in
this paper is defined to be%
\begin{equation}
|z|_{c}^{2}=z^{\circledast}z=(z_{0})^{2}+(z_{1})^{2}+(z_{2})^{2}+(z_{3})^{2}%
\end{equation}
which is a \textit{complex} number in general as a subscript $c$ is used in
the norm.

We are now ready to proceed the construction of $SL(2,C)$ instantons.
Historically, the general procedure to construct ADHM $Sp(N)$, $SU(N)$ and
$O(N)$ instantons are similar \cite{CSW}. The construction strongly relied on
the quaternion calculation. In this section, instead of $SU(2)$, we will
extend the $Sp(1)$ quaternion construction to the $SL(2,C)$ biquaternion
construction. We begin by introducing the $(k+1)\times k$ biquarternion matrix
$\Delta(x)=a+bx$%

\begin{equation}
\Delta(x)_{ab}=a_{ab}+b_{ab}x,\text{ }a_{ab}=a_{ab}^{\mu}e_{\mu},b_{ab}%
=b_{ab}^{\mu}e_{\mu} \label{ab}%
\end{equation}
where $a_{ab}^{\mu}$ and $b_{ab}^{\mu}$ are complex numbers, and $a_{ab}$ and
$b_{ab}$ are biquarternions. The biconjugation of the $\Delta(x)$ matrix is
defined to be%
\begin{equation}
\Delta(x)_{ab}^{\circledast}=\Delta(x)_{ba}^{\mu}e_{\mu}^{\dagger}%
=\Delta(x)_{ba}^{0}e_{0}-\Delta(x)_{ba}^{1}e_{1}-\Delta(x)_{ba}^{2}%
e_{2}-\Delta(x)_{ba}^{3}e_{3}.
\end{equation}

In contrast to the of $SU(2)$ instantons, the quadratic condition of $SL(2,C)$
instantons reads%

\begin{equation}
\Delta(x)^{\circledast}\Delta(x)=f^{-1}=\text{symmetric, non-singular }k\times
k\text{ matrix for }x\notin J\text{,} \label{ff}%
\end{equation}
from which\ we can deduce that $a^{\circledast}a,b^{\circledast}%
a,a^{\circledast}b$ and $b^{\circledast}b$ are all symmetric matrices. We
stress here that it will turn out the choice of \textit{biconjugation}
operation is crucial for the follow-up discussion in this paper. On the other
hand, for $x\in J,$ $\det\Delta(x)^{\circledast}\Delta(x)=0$. The set $J$ is
called singular locus or "jumping lines" in the mathematical literatures and
will be discussed in section IV.D. The existence of jumping lines is quite
common in complex ADHM equations. In contrast to the $SL(2,C)$ instantons,
there are no jumping lines for the case of $SU(2)$ instantons. We will assume
$x\notin J$ in the discussion for the rest of this section.

In the $Sp(1)$ quaternion case, the symmetric condition on $f^{-1}$ means
$f^{-1}$ is real. For the $SL(2,C)$ biquaternion case, however, it can be
shown that symmetric condition on $f^{-1}$ implies $f^{-1}$ is
\textit{complex}. Indeed, since%

\begin{align}
&  [\Delta(x)^{\circledast}\Delta(x)]_{ij}=\overset{k+1}{\underset{m=1}{\sum}%
}[\Delta(x)^{\circledast}]_{im}[\Delta(x)]_{mj}\nonumber\\
&  =\overset{k+1}{\underset{m=1}{\sum}}([\Delta(x)]_{mi}^{\mu}[\Delta
(x)]_{mj}^{\nu})(e_{\mu}^{\dagger}e_{\nu})=\overset{k+1}{\underset{m=1}{\sum}%
}([\Delta(x)]_{mj}^{\nu}[\Delta(x)]_{mi}^{\mu})(e_{\nu}^{\dagger}e_{\mu
})^{\dagger}\nonumber\\
&  =\overset{k+1}{\underset{m=1}{\sum}}\{([\Delta(x)]_{jm}^{\nu}e_{\nu
}^{\dagger})^{\circledast}([\Delta(x)]_{mi}^{\mu}e_{\mu})\}^{\circledast
}=[\Delta(x)^{\circledast}\Delta(x)]_{ji}^{\circledast},
\end{align}
the symmetric condition implies%

\begin{equation}
\lbrack\Delta(x)^{\circledast}\Delta(x)]_{ij}=[\Delta(x)^{\circledast}%
\Delta(x)]_{ij}^{\circledast},
\end{equation}
which means%

\begin{equation}
\lbrack\Delta(x)^{\circledast}\Delta(x)]_{ij}^{\mu}e_{\mu}=[\Delta
(x)^{\circledast}\Delta(x)]_{ij}^{\mu}e_{\mu}^{\dagger}.
\end{equation}
Thus only $[\Delta(x)^{\circledast}\Delta(x)]_{ij}^{0}$ is nonvanishing, and
it is in general a complex number for the case of biquaternion.

To construct the self-dual gauge field, we introduce a $(k+1)\times1$
dimensional biquaternion vector $v(x)$ satisfying the following two conditions%
\begin{align}
v^{\circledast}(x)\Delta(x)  &  =0,\label{null}\\
v^{\circledast}(x)v(x)  &  =1. \label{norm2}%
\end{align}
Note that $v(x)$ is fixed up to a $SL(2,C)$ gauge transformation%
\begin{equation}
v(x)\longrightarrow v(x)g(x),\text{ \ \ }g(x)\in\text{ }1\times1\text{
Biquaternion}.
\end{equation}
Note that in general a $SL(2,C)$ matrix can be written in terms of a
$1\times1$ biquaternion as%
\begin{equation}
g=\frac{q_{\mu}e_{\mu}}{\sqrt{q^{\circledast}q}}=\frac{q_{\mu}e_{\mu}}%
{|q|_{c}}.
\end{equation}
It is obvious that Eq.(\ref{null}) and Eq.(\ref{norm2}) are invariant under
the gauge transformation. The next step is to define the gauge field%

\begin{equation}
G_{\mu}(x)=v^{\circledast}(x)\partial_{\mu}v(x), \label{A}%
\end{equation}
which is a $1\times1$ biquaternion. The $SL(2,C)$ gauge transformation of the
gauge field is%

\begin{align}
G_{\mu}(x)-  &  >G^{\prime}(x)=(g^{\circledast}(x)v^{\circledast}%
(x))\partial_{\mu}(v(x)g(x))\nonumber\\
&  =g^{\circledast}(x)G_{\mu}(x)g(x)+g^{\circledast}(x)\partial_{\mu}g(x)
\end{align}
where in the calculation Eq.(\ref{norm2}) has been used. Note that, unlike the
case for $Sp(1)$, $G_{\mu}(x)$ needs not to be anti-Hermitian.

We can now define the $SL(2,C)$ field strength
\begin{equation}
F_{\mu\nu}=\partial_{\mu}G_{\nu}(x)+G_{\mu}(x)G_{\nu}(x)-[\mu
\longleftrightarrow\nu].
\end{equation}
To show that $F_{\mu\nu}$ is self-dual, one needs to show that the operator
\begin{equation}
P=1-v(x)v^{\circledast}(x)
\end{equation}
is a projection operator $P^{2}=P$, and can be written in terms of $\Delta$ as%

\begin{equation}
P=\Delta(x)f\Delta^{\circledast}(x). \label{P}%
\end{equation}
In fact%

\begin{align}
P^{2}  &  =(1-v(x)v^{\circledast}(x))(1-v(x)v^{\circledast}(x))\nonumber\\
&  =1-2v(x)v^{\circledast}(x)+v(x)v^{\circledast}(x)v(x)v^{\circledast
}(x)\nonumber\\
&  =1-v(x)v^{\circledast}(x)=P,
\end{align}
and%

\begin{equation}
Pv(x)=(1-v(x)v^{\circledast}(x))v(x)=v(x)-v(x)v^{\circledast}(x)v(x)=0.
\end{equation}
On the other hand%
\begin{equation}
P_{2}\equiv\Delta(x)f\Delta^{\circledast}(x),
\end{equation}

\begin{equation}
P_{2}^{2}=\Delta(x)f\Delta^{\circledast}(x)\Delta(x)f\Delta^{\circledast
}(x)=\Delta(x)ff^{-1}f\Delta^{\circledast}(x)=\Delta(x)f\Delta^{\circledast
}(x)=P_{2},
\end{equation}
and%
\begin{equation}
P_{2}v(x)=\Delta(x)f\Delta^{\circledast}(x)v(x)=0.
\end{equation}
So $P_{2}=P.$ This completes the proof. The self-duality of $F_{\mu\nu}$ can
now be proved as following

$\bigskip$%
\begin{align}
F_{\mu\nu}  &  =\partial_{\mu}(v^{\circledast}(x)\partial_{\nu}%
v(x))+v^{\circledast}(x)\partial_{\mu}v(x)v^{\circledast}(x)\partial_{\nu
}v(x)-[\mu\longleftrightarrow\nu]\nonumber\\
&  =\partial_{\mu}v^{\circledast}(x)[1-v(x)v^{\circledast}(x)]\partial_{\nu
}v(x)-[\mu\longleftrightarrow\nu]\nonumber\\
&  =\partial_{\mu}v^{\circledast}(x)\Delta(x)f\Delta^{\circledast}%
(x)\partial_{\nu}v(x)-[\mu\longleftrightarrow\nu]\nonumber\\
&  =v^{\circledast}(x)(\partial_{\mu}\Delta(x))f(\partial_{\nu}\Delta
^{\circledast}(x))v(x)-[\mu\longleftrightarrow\nu]\nonumber\\
&  =v^{\circledast}(x)(be_{\mu})f(e_{\nu}^{\dagger}b^{\circledast}%
)v(x)-[\mu\longleftrightarrow\nu]\nonumber\\
&  =v^{\circledast}(x)b(e_{\mu}e_{\nu}^{\dagger}-e_{\nu}e_{\mu}^{\dagger
})fb^{\circledast}v(x) \label{F}%
\end{align}
where we have used Eqs.(\ref{ab}),(\ref{null}) and (\ref{P}). Finally the
factor $(e_{\mu}e_{\nu}^{\dagger}-e_{\nu}e_{\mu}^{\dagger})$ above can be
shown to be self-dual%
\begin{align}
\sigma_{\mu\nu}  &  \equiv\frac{1}{4i}(e_{\mu}e_{\nu}^{\dagger}-e_{\nu}e_{\mu
}^{\dagger})=\frac{1}{2}\epsilon_{\mu\nu\alpha\beta}\sigma_{\alpha\beta
},\label{duall}\\
\overset{\_}{\sigma}_{\mu\nu}  &  =\frac{1}{4i}(e_{\mu}^{\dagger}e_{\nu
}-e_{\nu}^{\dagger}e_{\mu})=-\frac{1}{2}\epsilon_{\mu\nu\alpha\beta
}\overset{\_}{\sigma}_{\alpha\beta}.
\end{align}
This proves the self-duality of $F_{\mu\nu}.$ We thus have constructed many
$SL(2,C)$ SDYM field configurations.

To count the number of moduli parameters for the $SL(2,C)$ $k$-instantons we
have constructed , we will use transformations which preserve conditions
Eq.(\ref{ff}), Eq.(\ref{null}) and Eq.(\ref{norm2}), and the definition of
$G_{\mu}$ in Eq.(\ref{A}) to bring $a$ and $b$ in Eq.(\ref{ab}) into a simple
canonical form. The allowed transformations are similar to the case of $Sp(1)$
except that for the $SL(2,C)$ case$,$ $Q$ is unitary biquaternionic and
$K^{\circledast}=K^{T}$. That is%

\begin{equation}
a\rightarrow QaK,b\rightarrow QbK,v\rightarrow Qv
\end{equation}
where%

\begin{equation}
Q:(k+1)\times(k+1),\text{ }Q^{\circledast}Q=I\text{ },
\end{equation}

\begin{equation}
K^{\circledast}=K^{T}.
\end{equation}
One can use $K$ and $Q$ to bring $b$ to the following form%

\begin{equation}
b=%
\begin{bmatrix}
0_{1\times k}\\
I_{k\times k}%
\end{bmatrix}
\label{b}%
\end{equation}
Now the form of $b$ above is preserved by the following transformations

\bigskip%
\begin{equation}
Q=%
\begin{bmatrix}
Q_{1\times1} & 0\\
0 & X
\end{bmatrix}
,K=X^{T},Q_{1\times1}\in SL(2,C),X\in O(k).
\end{equation}
Then by choosing $X$ appropriately, one can diagonalize $a^{\circledast}a$ and
bring $a$ to the following form%

\begin{equation}
a=%
\begin{bmatrix}
\lambda_{1\times k}\\
-y_{k\times k}%
\end{bmatrix}
\label{a}%
\end{equation}
where $\lambda$ and $y$ are biquaternion matrices with orders $1\times k$ and
\ $k\times k$ respectively, and $y$ is symmetric%

\begin{equation}
y=y^{T}.
\end{equation}
Thus the constraints for the moduli parameters are%
\begin{equation}
a_{ci}^{\circledast}a_{cj}=0,i\neq j,\text{ and \ }y_{ij}=y_{ji}. \label{dof}%
\end{equation}
The forms $a$ and $b$ in Eq.(\ref{a}) and Eq.(\ref{b}) are called the
canonical forms of the construction, and $\lambda_{1\times k}$ , $y_{k\times
k}$ under the constraints Eq.(\ref{dof}) are the moduli parameters of
$k$-instantons. The total number of moduli parameters for $k$-instanton can be
calculated through Eq.(\ref{dof}) to be%
\begin{equation}
\text{\# of moduli for }SL(2,C)\text{ }k\text{-instantons}=16k-6,
\end{equation}
which is twice of that of the case of $Sp(1).$ Roughly speaking, there are
$8k$ parameters for instanton "biquaternion positions" and $8k$ parameters for
instanton "sizes". Finally one has to subtract an overall $SL(2,C)$ gauge
group degree of freesom $6.$ This picture will become more clear when we give
examples of explicit constructions of $SL(2,C)$ instantons in the next section.

\section{Examples of $SL(2,C)$ ADHM Instantons}

In this section, we will explicitly construct three examples of $SL(2,C)$ YM
instantons to illustrate our prescription given in the last section. More
importantly, we will also discuss the singular structures of $SL(2,C)$
$k$-instantons and compare our results with those in the mathematical literatures.

\subsection{The $SL(2,C)$ $(M,N)$ Instantons in ADHM Construction}

In this first example, we will reproduce from the ADHM construction the
$SL(2,C)$ $(M,N)$ instanton solutions \cite{Lee} discussed in section II. We
choose the biquaternion $\lambda_{j}$ in Eq.(\ref{a}) to be $\lambda_{j}e_{0}$
with $\lambda_{j}$ a \textit{complex} number, and choose $y_{ij}=y_{j}%
\delta_{ij}$ to be a diagonal matrix with $y_{j}=y_{j\mu}e_{\mu}$ a
quaternion. That is%

\begin{equation}
\Delta(x)=%
\begin{bmatrix}
\lambda_{1} & \lambda_{2} & ... & \lambda_{k}\\
x-y_{1} & 0 & ... & 0\\
0 & x-y_{2} & ... & 0\\
. & ... & ... & ...\\
0 & 0 & ... & x-y_{k}%
\end{bmatrix}
, \label{delta}%
\end{equation}
which satisfies the constraint in Eq.(\ref{dof}). Let%
\begin{equation}
v=\frac{1}{\sqrt{\phi}}%
\begin{bmatrix}
1\\
-q_{1}\\
.\\
-q_{k}%
\end{bmatrix}
,
\end{equation}
then
\begin{equation}
q_{j}=\frac{\lambda_{j}(x_{\mu}-y_{j\mu})e_{\mu}}{|x-y_{j}|^{2}},j=1,2,...,k,
\end{equation}
and%

\begin{equation}
v=\frac{1}{\sqrt{\phi}}%
\begin{bmatrix}
1\\
-\frac{\lambda_{1}(x_{\mu}-y_{1\mu})e_{\mu}}{|x-y_{1}|^{2}}\\
.\\
-\frac{\lambda_{k}(x_{\mu}-y_{k\mu})e_{\mu}}{|x-y_{k}|^{2}}%
\end{bmatrix}
\end{equation}
with

\bigskip%
\begin{equation}
\phi=1+\frac{\lambda_{1}^{2}}{|x-y_{1}|^{2}}+...+\frac{\lambda_{k}^{2}%
}{|x-y_{k}|^{2}}. \label{fai}%
\end{equation}
We have used $\lambda_{j}\lambda_{j}^{\circledast}=\lambda_{j}^{2}$ where
$\lambda_{j}^{2}$ a complex number in the above calculation. For the case of
$Sp(1),$ $\lambda_{j}$ is a real number and $\lambda_{j}\lambda_{j}^{\dagger
}=\lambda_{j}^{2}$ is a real number. So $\phi$ in Eq.(\ref{fai}) is a
complex-valued function in general. One can calculate the gauge potential as%
\begin{align}
G_{\mu}  &  =v^{\circledast}\partial_{\mu}v=\frac{1}{4}[e_{\mu}^{\dagger
}e_{\nu}-e_{\nu}^{\dagger}e_{\mu}]\partial_{\nu}\ln(1+\frac{\lambda_{1}^{2}%
}{|x-y_{1}|^{2}}+...+\frac{\lambda_{k}^{2}}{|x-y_{k}|^{2}})\nonumber\\
&  =\frac{1}{4}[e_{\mu}^{\dagger}e_{\nu}-e_{\nu}^{\dagger}e_{\mu}%
]\partial_{\nu}\ln(\phi).
\end{align}
If we choose $k=1$ and define $\lambda_{1}^{2}=\frac{\alpha_{1}^{2}}{1+i},$
then%
\begin{equation}
\phi=1+\frac{\frac{\alpha_{1}^{2}}{1+i}}{|x-y_{1}|^{2}}.
\end{equation}
The gauge potential is%
\begin{align}
G_{\mu}  &  =\frac{1}{4}[e_{\mu}^{\dagger}e_{\nu}-e_{\nu}^{\dagger}e_{\mu
}]\partial_{\nu}\ln(1+\frac{\frac{\alpha_{1}^{2}}{1+i}}{|x-y_{1}|^{2}}%
)=\frac{1}{4}[e_{\mu}^{\dagger}e_{\nu}-e_{\nu}^{\dagger}e_{\mu}]\partial_{\nu
}\ln(1+\frac{\alpha_{1}^{2}}{|x-y_{1}|^{2}}+i)\nonumber\\
&  =\frac{1}{2}[e_{\mu}^{\dagger}e_{\nu}-e_{\nu}^{\dagger}e_{\mu}%
]\frac{-\alpha_{1}^{2}(x-y_{1})_{\nu}}{|x-y_{1}|^{4}+(|x-y_{1}|^{2}+\alpha
_{1}^{2})^{2}}[\frac{|x-y_{1}|^{2}+\alpha_{1}^{2}}{|x-y_{1}|^{2}}-i]
\end{align}
which reproduces the $SL(2,C)$ $(M,N)=(1,0)$ solution calculated in
Eq.(\ref{(1,0)}). If we choose $k=1$ and consider $\lambda_{1}^{2}%
=\frac{i\beta_{1}^{2}}{1+i},$ then%

\begin{equation}
\phi=1+\frac{\frac{i\beta_{1}^{2}}{1+i}}{|x-y_{1}|^{2}}.
\end{equation}
The gauge potential is%
\begin{equation}
G_{\mu}=\frac{1}{4}[e_{\mu}^{\dagger}e_{\nu}-e_{\nu}^{\dagger}e_{\mu}%
]\partial_{\nu}\ln(1+\frac{\frac{i\beta_{1}^{2}}{1+i}}{|x-y_{1}|^{2}}%
)=\frac{1}{4}[e_{\mu}^{\dagger}e_{\nu}-e_{\nu}^{\dagger}e_{\mu}]\partial_{\nu
}\ln[(1+i+\frac{i\beta_{1}^{2}}{|x-y_{1}|^{2}})],
\end{equation}
which reproduces the $SL(2,C)$ $(M,N)=(0,1)$ solution calculated in
Eq.(\ref{(0,1)}). If we choose $k=2$ and $\lambda_{1}^{2}=\frac{\alpha_{1}%
^{2}}{1+i},\lambda_{2}^{2}=\frac{i\beta_{1}^{2}}{1+i},$ we get%
\begin{align}
\phi &  =1+\frac{\frac{\alpha_{1}^{2}}{1+i}}{|x-y_{1}|^{2}}+\frac{\frac
{i\beta_{1}^{2}}{1+i}}{|x-y_{2}|^{2}},\\
G_{\mu}  &  =\frac{1}{4}[e_{\mu}^{\dagger}e_{\nu}-e_{\nu}^{\dagger}e_{\mu
}]\partial_{\nu}\ln(1+\frac{\frac{\alpha_{1}^{2}}{1+i}}{|x-y_{1}|^{2}}%
+\frac{\frac{i\beta_{1}^{2}}{1+i}}{|x-y_{2}|^{2}})\\
&  =\frac{1}{4}[e_{\mu}^{\dagger}e_{\nu}-e_{\nu}^{\dagger}e_{\mu}%
]\partial_{\nu}\ln(1+i+\frac{\alpha_{1}^{2}}{|x-y_{1}|^{2}}+\frac{i\beta
_{1}^{2}}{|x-y_{2}|^{2}}),
\end{align}
which reproduces the $SL(2,C)$ $(1,1)$ solution calculated in Eq.(\ref{(1,1)}%
). It is easy to generalize the above calculations to the general $(M,N)$
cases. The $SL(2,C)$ ADHM $k$-instanton solutions we proposed in section III
thus include the $SL(2,C)$ $(M,N)$ $k$-instanton solutions calculated
previously in \cite{Lee} as a subset.

\subsection{The $SL(2,C)$ $k=2,3$ Instanton Solutions}

For the case of $2$-instantons, we begin with the following $\Delta(x)$ matrix
with $y_{12}=y_{21}$%

\begin{equation}
\Delta(x)=%
\begin{bmatrix}
\lambda_{1} & \lambda_{2}\\
x-y_{1} & -y_{12}\\
-y_{21} & x-y_{2}%
\end{bmatrix}
, \label{y12}%
\end{equation}

\begin{equation}
\Delta^{\circledast}(x)=%
\begin{bmatrix}
\lambda_{1}^{\circledast} & x^{\circledast}-y_{1}^{\circledast} &
-y_{12}^{\circledast}\\
\lambda_{2}^{\circledast} & -y_{12}^{\circledast} & x^{\circledast}%
-y_{2}^{\circledast}%
\end{bmatrix}
.
\end{equation}
The condition on $\Delta^{\circledast}(x)\Delta(x)$%
\begin{equation}
\Delta^{\circledast}(x)\Delta(x)=%
\begin{bmatrix}
\lambda_{1}^{\circledast}\lambda_{1}+(x^{\circledast}-y_{1}^{\circledast
})(x-y_{1})+y_{12}^{\circledast}y_{12} & \lambda_{1}^{\circledast}\lambda
_{2}-(x^{\circledast}-y_{1}^{\circledast})y_{12}-y_{12}^{\circledast}%
(x-y_{2})\\
\lambda_{2}^{\circledast}\lambda_{1}-y_{12}^{\circledast}(x-y_{1}%
)-(x^{\circledast}-y_{2}^{\circledast})y_{12} & \lambda_{2}^{\circledast
}\lambda_{2}+y_{12}^{\circledast}y_{12}+(x^{\circledast}-y_{2}^{\circledast
})(x-y_{2})
\end{bmatrix}
\label{2-instan}%
\end{equation}
in Eq.(\ref{ff}) is\bigskip%

\begin{equation}
\lambda_{2}^{\circledast}\lambda_{1}-\lambda_{1}^{\circledast}\lambda
_{2}=y_{12}^{\circledast}(y_{2}-y_{1})+(y_{1}^{\circledast}-y_{2}%
^{\circledast})y_{12},
\end{equation}
which is linear in the biquaternion $y_{12}$ instead of a quadratic equation,
and $y_{12}$ can be easily solved to be%
\begin{equation}
y_{12}=\frac{1}{2}\frac{(y_{1}-y_{2})}{|y_{1}-y_{2}|_{c}^{2}}(\lambda
_{2}^{\circledast}\lambda_{1}-\lambda_{1}^{\circledast}\lambda_{2}).
\label{yy}%
\end{equation}
So the four biquaternions $y_{1},y_{2},\lambda_{1}$ and $\lambda_{2}$ gives
$4\times8=32$ real parameters. After subtracting $SL(2,C)$ gauge group degree
of freesom $6,$ the number of moduli for $SL(2,C)$ $2$-instanton is $26$ as
expected. The result in Eq.(\ref{yy}) is the same with the case of $Sp(1)$
except with quaternions replaced by biquaternions \cite{CSW}. However, for the
$SL(2,C)$ $2$-instantons in constrast to the $SU(2)$ case, there are so-called
jumping lines in zeros of determinant of Eq.(\ref{2-instan}) which correspond
to singularities of $SL(2,C)$ $2$-instanton field configurations. This will be
calculated in section IV. D.

For the case of $3$-instantons, we begin with the following $\Delta(x)$ matrix
with $y_{ij}=y_{ji}$%

\begin{equation}
\Delta(x)=%
\begin{bmatrix}
\lambda_{1} & \lambda_{2} & \lambda_{3}\\
x-y_{1} & -y_{12} & -y_{13}\\
-y_{21} & x-y_{2} & -y_{23}\\
-y_{31} & -y_{32} & x-y_{3}%
\end{bmatrix}
. \label{3-ins}%
\end{equation}
In order to get the general solutions for $k=3$ $SL(2,C)$ instanton, similar
to \cite{CSW}, we make the choices $\lambda_{1}=\lambda_{1}^{0}\otimes e_{0}$
($\lambda_{1}^{1}=\lambda_{1}^{2}=\lambda_{1}^{3}=0$) and $y_{12}^{0}%
=y_{13}^{0}=y_{23}^{0}=0$. Then the remaining parameters are the positions
$y_{1},y_{2},y_{3}$ and the imaginary part of $y_{12},y_{13},y_{23}$. So there
are $8\times3+6\times3=42=16k-6(k=3)$ parameters. Other parameters can be
fixed by constraints to be%
\begin{align}
\lambda_{1}  &  =\lambda_{1}^{0}\otimes e_{0}\label{1}\\
\lambda_{1}^{0}  &  =\frac{|\overrightarrow{W_{2}}\times\overrightarrow{W_{3}%
}|_{c}}{|\overrightarrow{W_{1}}\cdot(\overrightarrow{W_{2}}\times
\overrightarrow{W_{3}})|_{c}^{1/2}}\label{2}\\
\lambda_{2}  &  =\lambda_{1}\frac{(\overrightarrow{W_{3}}\times
\overrightarrow{W_{2}})\cdot(\overrightarrow{W_{3}}\times\overrightarrow{W_{1}%
})}{|\overrightarrow{W_{2}}\times\overrightarrow{W_{3}}|_{c}^{2}%
}+i\overrightarrow{\sigma}\cdot\frac{1}{\lambda_{1}}\overrightarrow{W_{3}%
}\label{3}\\
\lambda_{3}  &  =\lambda_{1}\frac{(\overrightarrow{W_{3}}\times
\overrightarrow{W_{2}})\cdot(\overrightarrow{W_{2}}\times\overrightarrow{W_{1}%
})}{|\overrightarrow{W_{2}}\times\overrightarrow{W_{3}}|_{c}^{2}%
}-i\overrightarrow{\sigma}\cdot\frac{1}{\lambda_{1}}\overrightarrow{W_{2}}
\label{4}%
\end{align}
where the vectors $\overrightarrow{W_{k}}$ are defined by%
\begin{equation}
\overrightarrow{W_{k}}=\frac{i}{4}\epsilon_{ijk}tr\{\overrightarrow{\sigma
}[(y_{i}-y_{j})^{\circledast}y_{ij}+\underset{l=1}{\overset{3}{\sum}}%
(y_{li}^{\circledast}y_{lj})]\}. \label{5}%
\end{equation}
Here we have presented the biquaternions $\lambda_{i}$ as $2\times2$ matrices.
The result in the above equations are the same with the case of $Sp(1)$
\cite{CSW} except with quaternions replaced by biquaternions. However, for the
$SL(2,C)$ $3$-instantons in constrast to the $SU(2)$ case, there are jumping
lines of $3$-instantons corresponding to Eq.(\ref{3-ins}). This will be
discussed in section IV. D.

\subsection{The $SL(2,C)$ $1$-Instanton Solution and its Singularities}

In the third example, we calculate the complete $SL(2,C)$ $10$ parameters
$1$-instanton solution and study structure of its singularities. The singular
structures of a subset of $SL(2,C)$ $k$-instantons will be discussed in the
next subsection. We will see that the singularities for $SL(2,C)$
$1$-instanton is much more complicated that that of $SU(2)$ $1$-instanton. All
$10$ parameters are closely related to the structure of the singularities. We
first build $\Delta(x)$ matrix and choose $a,b$ as%
\begin{equation}
a=%
\begin{bmatrix}
\lambda\\
-y
\end{bmatrix}
,b=%
\begin{bmatrix}
0\\
1
\end{bmatrix}
,
\end{equation}

\begin{equation}
\Delta(x)=a+bx=%
\begin{bmatrix}
\lambda\\
x-y
\end{bmatrix}
\end{equation}
where $x$ is a quaternion, $\lambda=\lambda e_{0}$ (with $\lambda$ a
\textit{complex }number) and$\ y$ is a biquaternion. It can be checked that,
for these choices, the constraints in Eq.(\ref{dof}) are satisfied. By
Eq.(\ref{null}) and Eq.(\ref{norm2}), we easily obtain%

\begin{equation}
v(x)=\frac{1}{\sqrt{\phi}}%
\begin{bmatrix}
1\\
-\frac{(x-y)\lambda^{\circledast}}{|x-y|_{c}^{2}}%
\end{bmatrix}
\end{equation}
with%

\begin{equation}
\phi=1+\frac{\lambda\lambda^{\circledast}}{|x-y|_{c}^{2}}. \label{one}%
\end{equation}
Note that $\lambda\lambda^{\circledast}=\lambda^{2}$ is a complex number and
$|x-y|_{c}^{2}\equiv|x-(p+qi)|_{c}^{2}$ is also a complex number. Here $p$ and
$q$ are quaternions. The total number of moduli parameters is thus $10.$ The
gauge field $G_{\mu}$ can be calculated to be%

\begin{align}
G_{\mu}  &  =v^{\circledast}\partial_{\mu}v(x)\nonumber\\
&  =\frac{1}{4}[e_{\mu}^{\dagger}e_{\nu}-e_{\nu}^{\dagger}e_{\mu}%
]\partial_{\nu}\ln(1+\frac{\lambda^{2}}{|x-y|_{c}^{2}})\nonumber\\
&  =\frac{-1}{2}[e_{\mu}^{\dagger}e_{\nu}-e_{\nu}^{\dagger}e_{\mu}%
]\frac{[x-(p+qi)]_{\nu}\lambda^{2}}{|x-(p+qi)|_{c}^{2}(|x-(p+qi)|_{c}%
^{2}+\lambda^{2})}. \label{GG}%
\end{align}
By solving $|x-(p+qi)|_{c}^{2}=0$ in the denominator of Eq.(\ref{GG}), we can
get some singularities of $G_{\mu}.$ We see that%
\begin{align}
&  |x-(p+qi)|_{c}^{2}=[(x_{0}-p_{0})^{2}+(x_{1}-p_{1})^{2}+(x_{2}-p_{2}%
)^{2}+(x_{3}-p_{3})^{2})-(q_{0}^{2}+q_{1}^{2}+q_{2}^{2}+q_{3}^{2})]\nonumber\\
&  -2i[(x_{0}-p_{0})q_{0}+(x_{1}-p_{1})q_{1}+(x_{2}-p_{2})q_{2}+(x_{3}%
-p_{3})q_{3}]=0
\end{align}
implies

\bigskip\
\begin{align}
\lbrack(x_{0}-p_{0})^{2}+(x_{1}-p_{1})^{2}+(x_{2}-p_{2})^{2}+(x_{3}-p_{3}%
)^{2}]  &  =(q_{0}^{2}+q_{1}^{2}+q_{2}^{2}+q_{3}^{2}),\label{p}\\
(x_{0}-p_{0})q_{0}+(x_{1}-p_{1})q_{1}+(x_{2}-p_{2})q_{2}+(x_{3}-p_{3})q_{3}
&  =0. \label{q}%
\end{align}
Generically for $q\neq0$, Eq.(\ref{p}) and Eq.(\ref{q}) describe in $R^{4}$ a
$S^{3}$ and an hyper-plane $R^{3}$ passing through the center of the $S^{3}$
respectively. Thus the intersection of these $S^{3}$ and $R^{3}$ is a $S^{2}$.
This means that the singularities is a $S^{2}$ in $R^{4}$. It is clear
geometrically that $p_{\mu}$ is the center of the $S^{2}$ and $q_{\mu}$ gives
radius and orientation of the $S^{2}$ in $R^{4}$. In fact, these singularities
can be gauged away just like in the $SU(2)$ case. If we define%

\begin{equation}
U_{1c}(z)=\frac{z}{|z|_{c}}=\frac{(x-p-qi)^{\mu}e_{\mu}}{|x-p-qi|_{c}}%
\end{equation}
where $U_{1c}(z)$ is a $1\times1$ biquaternion corresponding to a $SL(2,C)$
matrix, which is to be compared with Eq.(\ref{SU}) for the case of $SU(2)$. Then

\bigskip%
\begin{equation}
U_{1c}(z)\frac{\partial}{\partial z^{\mu}}U_{1c}^{-1}(z)=\frac{z}{|z|_{c}%
}\frac{\partial}{\partial z^{\mu}}\frac{z^{\circledast}}{|z|_{c}}=-\frac{1}%
{2}[e_{\mu}e_{\nu}^{\dagger}-e_{\nu}e_{\mu}^{\dagger}]\frac{z_{\nu}}%
{|z|_{c}^{2}},
\end{equation}%
\begin{equation}
U_{1c}^{-1}(z)\frac{\partial}{\partial z^{\mu}}U_{1c}(z)=\frac{z^{\circledast
}}{|z|_{c}}\frac{\partial}{\partial z^{\mu}}\frac{z}{|z|_{c}}=-\frac{1}%
{2}[e_{\mu}^{\dagger}e_{\nu}-e_{\nu}^{\dagger}e_{\mu}]\frac{z_{\nu}}%
{|z|_{c}^{2}}.
\end{equation}
It's easy to see that $G_{\mu}$ can be written as

\bigskip%
\begin{align}
G_{\mu}  &  =\frac{1}{2}[e_{\mu}^{\dagger}e_{\nu}-e_{\nu}^{\dagger}e_{\mu
}]\frac{-(x-(p+qi))_{\nu}\lambda^{2}}{|x-(p+qi)|_{c}^{2}(|x-(p+qi)|_{c}%
^{2}+\lambda^{2})}\nonumber\\
&  =U_{1c}^{-1}(z)\frac{\partial}{\partial z^{\mu}}U_{1c}(z)\frac{\lambda^{2}%
}{(|x-(p+qi)|_{c}^{2}+\lambda^{2})}.
\end{align}
We can now do the $SL(2,C)$ gauge transformation

\bigskip%

\begin{align}
G_{\mu}^{\prime}  &  =U_{1c}(z)G_{\mu}U_{1c}^{-1}(z)+U_{1c}(z)\frac{\partial
}{\partial z^{\mu}}U_{1c}^{-1}(z)\nonumber\\
&  =\frac{-1}{2}[e_{\mu}e_{\nu}^{\dagger}-e_{\nu}e_{\mu}^{\dagger}%
]\frac{[x-(p+qi)]_{\nu}}{(|x-(p+qi)|_{c}^{2}+\lambda^{2})}%
\end{align}
to gauge away the singularities caused by $|x-(p+qi)|_{c}^{2}=0$. But there
are still non-removable singularities remained, which come from
$(|x-(p+qi)|_{c}^{2}+\lambda^{2})=0$ in the denominator of Eq.(\ref{GG}). To
study these singularities, let the real part of $\lambda^{2}$ be $c$ and
imaginary part of $\lambda^{2}$ be $d,$ we see that%

\begin{align}
&  (|x-(p+qi)|_{c}^{2}+\lambda^{2})=P_{2}(x)+iP_{1}(x)\nonumber\\
&  =[(x_{0}-p_{0})^{2}+(x_{1}-p_{1})^{2}+(x_{2}-p_{2})^{2}+(x_{3}-p_{3}%
)^{2}-(q_{0}^{2}+q_{1}^{2}+q_{2}^{2}+q_{3}^{2})]+c\nonumber\\
&  -2i[(x_{0}-p_{0})q_{0}+(x_{1}-p_{1})q_{1}+(x_{2}-p_{2})q_{2}+(x_{3}%
-p_{3})q_{3}-\frac{d}{2}]=0 \label{jump1}%
\end{align}
implies%
\begin{align}
(x_{0}-p_{0})^{2}+(x_{1}-p_{1})^{2}+(x_{2}-p_{2})^{2}+(x_{3}-p_{3})^{2}  &
=(q_{0}^{2}+q_{1}^{2}+q_{2}^{2}+q_{3}^{2})-c,\label{p1}\\
(x_{0}-p_{0})q_{0}+(x_{1}-p_{1})q_{1}+(x_{2}-p_{2})q_{2}+(x_{3}-p_{3})q_{3}
&  =\frac{d}{2} \label{p2}%
\end{align}
where $P_{2}(x)$ and $P_{1}(x)$ are polynomials of $4$ variables with degree
$2$ and $1$ respectively.

For a subset of $k$-instanton field configurations, one chooses $\lambda
_{i}=\lambda_{i}e_{0}$ (with $\lambda_{i}$ a \textit{complex }number)
and$\ y_{i}$ to be a \textit{biquaternion} in Eq.(\ref{delta}). It is
important to note that for these choices, the constraints in Eq.(\ref{dof})
are still satisfied \textit{without} turning on the off-diagonal elements
$y_{ij}$ in Eq.(\ref{a}). To get non-removable singularities, one needs to
calculate zeros of%
\begin{equation}
\phi=1+\frac{\lambda_{1}\lambda_{1}^{\circledast}}{|x-y_{1}|_{c}^{2}%
}+...+\frac{\lambda_{k}\lambda_{k}^{\circledast}}{|x-y_{k}|_{c}^{2}},
\label{k}%
\end{equation}
or%
\begin{equation}
|x-y_{1}|_{c}^{2}|x-y_{2}|_{c}^{2}\cdot\cdot\cdot|x-y_{k}|_{c}^{2}\phi
=P_{2k}(x)+iP_{2k-1}(x)=0. \label{jump}%
\end{equation}
For the $k$-instanton case, one encounters intersections of zeros of
$P_{2k}(x)$ and $P_{2k-1}(x)$ polynomials with degrees $2k$ and $2k-1$
respectively%
\begin{equation}
P_{2k}(x)=0,\text{ \ }P_{2k-1}(x)=0. \label{pp}%
\end{equation}
We will discuss the singular structures of these $k$-instanton field
configurations in the next subsection.

The structure of singularities of $SL(2,C)$ $1$-instanton can be classified
into the following four cases:

$(1)\bigskip$ For $q=0$, Eq.(\ref{p2}) implies $d=0$. If $c<0$, one gets
$S^{3}$ singularities. If $c=0$, one gets a singular point at $p$.

The following three cases are for $q\neq0$.

$(2)$ If the $6$ parameters satisfy%
\begin{equation}
(q_{0}^{2}+q_{1}^{2}+q_{2}^{2}+q_{3}^{2})-c<\frac{\frac{d^{2}}{4}}{(q_{0}%
^{2}+q_{1}^{2}+q_{2}^{2}+q_{3}^{2})},
\end{equation}

\bigskip then there are no singularities.

\bigskip$(3)$ If the $6$ parameters satisfy%
\begin{equation}
(q_{0}^{2}+q_{1}^{2}+q_{2}^{2}+q_{3}^{2})-c=\frac{\frac{d^{2}}{4}}{(q_{0}%
^{2}+q_{1}^{2}+q_{2}^{2}+q_{3}^{2})},
\end{equation}

then there is only one singular point which is located at%

\begin{equation}
(x_{0},x_{1},x_{2},x_{3})=(p_{0},p_{1},p_{2},p_{3})+\frac{\frac{d}{2}}%
{(q_{0}^{2}+q_{1}^{2}+q_{2}^{2}+q_{3}^{2})}(q_{0},q_{1},q_{2},q_{3}).
\end{equation}

$(4)$ If the $6$ parameters satisfy%

\begin{equation}
(q_{0}^{2}+q_{1}^{2}+q_{2}^{2}+q_{3}^{2})-c>\frac{\frac{d^{2}}{4}}{(q_{0}%
^{2}+q_{1}^{2}+q_{2}^{2}+q_{3}^{2})}, \label{case4}%
\end{equation}

then the singularities are the intersection of a $R^{3}$ and a $S^{3}$, or a
$S^{2}$ surface, similar to the previous discussion in Eq.(\ref{p}) and
Eq.(\ref{q}). We can see that if $|q|$ is big enough, the $S^{2}$
singularities will be turned on in the $R^{4}$ space. Unlike singularities
which can be gauged away, it seems that these singularities can not be gauged away.

Finally the real parts and imaginary parts of the gauge field and the field
strength of $SL(2,C)$ $1$-instanton solution with $10$ moduli parameters can
be calculated to be%
\begin{align}
G_{\mu}^{\prime}  &  =\frac{-1}{2}[e_{\mu}e_{\nu}^{\dagger}-e_{\nu}e_{\mu
}^{\dagger}]\frac{[x-(p+qi)]_{\nu}}{(|x-(p+qi)|_{c}^{2}+\lambda^{2}%
)}\nonumber\\
&  =\frac{-1}{2}[e_{\mu}e_{\nu}^{\dagger}-e_{\nu}e_{\mu}^{\dagger}%
]\frac{\{[|x-p|^{2}-q^{2}+c](x-p)_{\nu}-[d-2(x-p)\cdot q]q_{\nu}\}}%
{[|x-p|^{2}-q^{2}+c]^{2}+[d-2(x-p)\cdot q]^{2}}\nonumber\\
&  -i\frac{1}{2}[e_{\mu}e_{\nu}^{\dagger}-e_{\nu}e_{\mu}^{\dagger}%
]\frac{\{-[|x-p|^{2}-p^{2}+c]q_{\nu}-[d-2(x-p)\cdot q](x-p)_{\nu}%
\}}{[|x-p|^{2}-q^{2}+c]^{2}+[d-2(x-p)\cdot q]^{2}}%
\end{align}
and%
\begin{align}
F_{\mu\nu}^{\prime}  &  =[e_{\mu}e_{\nu}^{\dagger}-e_{\nu}e_{\mu}^{\dagger
}]\frac{(c+di)}{[|x-(p+qi)|_{c}^{2}+(c+di)]^{2}}\nonumber\\
=  &  \frac{%
\begin{array}
[c]{c}%
\lbrack e_{\mu}e_{\nu}^{\dagger}-e_{\nu}e_{\mu}^{\dagger}]\{c[|x-p|^{2}%
-q^{2}+c]^{2}-c[d-2(x-p)\cdot q]^{2}\\
+2d[|x-p|^{2}-q^{2}+c][d-2(x-p)\cdot q]\}
\end{array}
}{\{[|x-p|^{2}-q^{2}+c]^{2}+[d-2(x-p)\cdot q]^{2}\}^{2}}\nonumber\\
&  +i\frac{%
\begin{array}
[c]{c}%
\lbrack e_{\mu}e_{\nu}^{\dagger}-e_{\nu}e_{\mu}^{\dagger}]\{d[|x-p|^{2}%
-q^{2}+c]^{2}-d[d-2(x-p)\cdot q]^{2}\\
-2c[|x-p|^{2}-q^{2}+c][d-2(x-p)\cdot q]\}
\end{array}
}{\{[|x-p|^{2}-q^{2}+c]^{2}+[d-2(x-p)\cdot q]^{2}\}^{2}}%
\end{align}
which is a self-dual field configuration by Eq.(\ref{duall}). If we take
$q=0,$ $c=\frac{\alpha_{1}^{2}}{2}=-d$, we can easily get the following
special solutions%

\begin{equation}
G_{\mu}^{\prime}=-\frac{1}{2}[e_{\mu}e_{\nu}^{\dagger}-e_{\nu}e_{\mu}%
^{\dagger}]\frac{[2|x-p|^{2}+\alpha_{1}^{2}+i\alpha_{1}^{2}](x-p)_{\nu}%
}{(|x-p|^{2})^{2}+2|x-p|^{2}\alpha_{1}^{2}+\alpha_{1}^{4}+|x-p|^{4}%
}\label{conn}%
\end{equation}
and%

\begin{align}
F_{\mu\nu}^{\prime}  &  =\frac{\alpha_{1}^{2}(e_{\mu}e_{\nu}^{\dagger}-e_{\nu
}e_{\mu}^{\dagger})[2|x-p|^{4}+4|x-p|^{2}\alpha_{1}^{2}+\alpha_{1}^{4}%
]}{[2|x-p|^{4}+2|x-p|^{2}\alpha_{1}^{2}+\alpha_{1}^{4}]^{2}}\nonumber\\
&  -i\frac{-\alpha_{1}^{2}(e_{\mu}e_{\nu}^{\dagger}-e_{\nu}e_{\mu}^{\dagger
})[2|x-p|^{4}-\alpha_{1}^{4}]}{[2|x-p|^{4}+2|x-p|^{2}\alpha_{1}^{2}+\alpha
_{1}^{4}]^{2}},
\end{align}
which can be written as
\begin{align}
A_{\mu}^{^{\prime}}  &  =-\frac{1}{4}[e_{\mu}e_{\nu}^{\dagger}-e_{\nu}e_{\mu
}^{\dagger}]y_{\nu}\frac{2(2y^{2}+\alpha_{1}^{2})}{y^{4}+(y^{2}+\alpha_{1}%
^{2})^{2}},\nonumber\\
B_{\mu}^{\prime}  &  =-\frac{1}{4}[e_{\mu}e_{\nu}^{\dagger}-e_{\nu}e_{\mu
}^{\dagger}]y_{\nu}\frac{2\alpha_{1}^{2}}{y^{4}+(y^{2}+\alpha_{1}^{2})^{2}},
\label{(1,0)L}%
\end{align}
and
\begin{align}
H_{\mu\nu}^{\prime}  &  =(e_{\mu}e_{\nu}^{\dagger}-e_{\nu}e_{\mu}^{\dagger
})\alpha_{1}^{2}\frac{2y^{4}+4\alpha_{1}^{2}y^{2}+\alpha_{1}^{4}}%
{[y^{4}+(y^{2}+\alpha_{1}^{2})^{2}]^{2}},\nonumber\\
M_{\mu\nu}^{\prime}  &  =(e_{\mu}e_{\nu}^{\dagger}-e_{\nu}e_{\mu}^{\dagger
})\alpha_{1}^{2}\frac{2y^{4}-\alpha_{1}^{4}}{[y^{4}+(y^{2}+\alpha_{1}^{2}%
)^{2}]^{2}} \label{(1,0)L2}%
\end{align}
where $y=x-p$, $y^{2}=|x-p|^{2}.$ The forms of profiles in Eqs.(\ref{(1,0)L})
and (\ref{(1,0)L2}) are exactly the same with the $(M,N)=(1,0)$ instanton
solution \cite{Lee} obtained in Eqs.(\ref{1,0}) and (\ref{1,0f}).

\subsection{The Jumping lines of $SL(2,C)$ $k$-Instantons}

For the subset of $SL(2,C)$ $k$-instantons in Eq.(\ref{k}), the connections
are calculable and one encounters much more complicated singular structures of
the field configurations in Eq.(\ref{pp}). These new singularities can not be
gauged away and do not show up in the field configurations of $SU(2)$
$k$-instantons. Mathematically, the existence of singular structures of the
non-compact $SL(2,C)$ SDYM field configurations is consistent with the
inclusion of "sheaves" by Frenkel-Jardim \cite{math2} recently, rather than
just the restricted notion of "vector bundles", in the one to one
correspondence between ASDYM and certain algebraic geometric objects.

In fact, one notices that Eq.(\ref{jump}) can be written as%
\begin{equation}
\det\Delta(x)^{\circledast}\Delta(x)=|x-y_{1}|_{c}^{2}|x-y_{2}|_{c}^{2}%
\cdot\cdot\cdot|x-y_{k}|_{c}^{2}\phi=P_{2k}(x)+iP_{2k-1}(x)=0 \label{k-jump}%
\end{equation}
where%
\begin{equation}
\Delta(x)=%
\begin{bmatrix}
\lambda_{1} & \lambda_{2} & ... & \lambda_{k}\\
x-y_{1} & 0 & ... & 0\\
0 & x-y_{2} & ... & 0\\
. & ... & ... & ...\\
0 & 0 & ... & x-y_{k}%
\end{bmatrix}
. \label{del}%
\end{equation}
In Eq.(\ref{del}) $\lambda_{i}=\lambda_{i}e_{0}$ (with $\lambda_{i}$ a
\textit{complex }number) and$\ y_{i}$ is a \textit{biquaternion }in contrast
to those in Eq.(\ref{delta}) where $y_{i}$ was chosen to be a
\textit{quaternion}. The solutions $x\in J$ of%
\begin{equation}
\det\Delta(x)^{\circledast}\Delta(x)=0
\end{equation}
mentioned in Eq.(\ref{ff}) are called "jumping lines" of $k$-instantons in the
mathematical literatures \cite{math2}.

The complete jumping lines of the $SL(2,C)$ $1$-instanton are given in
Eq.(\ref{jump1}). The jumping lines for the $SL(2,C)$ $k$-instantons given in
Eq.(\ref{k-jump}) with $k\geq2$ are just partial subsets of the jumping lines.
For the complete jumping lines of $2$-instantons, for example, $\lambda
_{1},\lambda_{2}$ and $y_{1},y_{2}$ are biquaternions and one needs to turn on
$y_{12}$ and $y_{21}(=y_{12})$ in Eq.(\ref{y12}) which are also biquaternions.
We can calculate the determinant of Eq.(\ref{2-instan}) in section IV. B to
get%
\begin{align}
\det\Delta_{2-ins}(x)^{\circledast}\Delta_{2-ins}(x) &  =|x-y_{1}|_{c}%
^{2}|x-y_{2}|_{c}^{2}+|\lambda_{2}|_{c}^{2}|x-y_{1}|_{c}^{2}+|\lambda_{1}%
|_{c}^{2}|x-y_{2}|_{c}^{2}\nonumber\\
&  +y_{12}^{\circledast}(x-y_{1})y_{12}^{\circledast}(x-y_{2})+(x-y_{2}%
)^{\circledast}y_{12}(x-y_{1})^{\circledast}y_{12}\nonumber\\
&  -y_{12}^{\circledast}(x-y_{1})\lambda_{1}^{\circledast}\lambda_{2}%
-\lambda_{2}^{\circledast}\lambda_{1}(x-y_{1})^{\circledast}y_{12}\nonumber\\
&  -(x-y_{2})^{\circledast}y_{12}\lambda_{1}^{\circledast}\lambda_{2}%
-\lambda_{2}^{\circledast}\lambda_{1}y_{12}^{\circledast}(x-y_{2})\nonumber\\
&  +|y_{12}|_{c}^{2}(|\lambda_{2}|_{c}^{2}+|\lambda_{1}|_{c}^{2})+|y_{12}%
|_{c}^{4}\nonumber\\
&  =0\label{jump2}%
\end{align}
where $y_{12}$ is given by Eq.(\ref{yy}). In calculating the determinant, one
notices that $\Delta(x)^{\circledast}\Delta(x)$ in Eq.(\ref{2-instan}) is a
symmetric matrix with complex number entries. So there is no ambiguity in the
determinant calculation. The result of the determinant in Eq.(\ref{jump2})
corresponds to a polynomial of degree $2k=4$ in the variable $x$, which is
consistent with results obtained in the mathematical literature \cite{math2}.
Here we have given the singularity locus or jumping lines in a much more
detailed form accessible for most physicists. For the special case of
$y_{12}=0$, $\lambda_{i}$ reduces to a complex number $\lambda_{i}=\lambda
_{i}e_{0}$ (with $\lambda_{i}$ a \textit{complex }number) and Eq.(\ref{jump2})
reduces to Eq.(\ref{k-jump}). It is interesting to see that although the
complete $SL(2,C)$ $2$-instanton connections with $26$ parameters are not
available, their jumping lines can be identified by Eq.(\ref{jump2}).

Similar results can be obtained for the jumping lines of the complete
$3$-instantons by using Eq.(\ref{3-ins}) although the calculation is more
involved%
\begin{align}
&  \det\Delta_{3-ins}(x)^{\circledast}\Delta_{3-ins}(x)\nonumber\\
&  =[|\lambda_{1}|_{c}^{2}+|x-y_{1}|_{c}^{2}+|y_{12}|_{c}^{2}+|y_{13}|_{c}%
^{2}]\nonumber\\
&  \times\lbrack|\lambda_{2}|_{c}^{2}+|x-y_{2}|_{c}^{2}+|y_{12}|_{c}%
^{2}+|y_{23}|_{c}^{2}]\nonumber\\
&  \times\lbrack|\lambda_{3}|_{c}^{2}+|x-y_{3}|_{c}^{2}+|y_{13}|_{c}%
^{2}+|y_{23}|_{c}^{2}]\nonumber\\
&  +[\lambda_{1}^{\circledast}\lambda_{2}-(x-y_{1})^{\circledast}y_{12}%
-y_{12}^{\circledast}(x-y_{2})+y_{13}^{\circledast}y_{23}]\nonumber\\
&  \times\lbrack\lambda_{2}^{\circledast}\lambda_{3}+y_{12}^{\circledast
}y_{13}-(x-y_{2})^{\circledast}y_{23}-y_{23}^{\circledast}(x-y_{3}%
)]\nonumber\\
&  \times\lbrack\lambda_{3}^{\circledast}\lambda_{1}-y_{13}^{\circledast
}(x-y_{1})+y_{23}^{\circledast}y_{12}-(x-y_{3})^{\circledast}y_{13}%
]\nonumber\\
&  +[\lambda_{1}^{\circledast}\lambda_{3}-(x-y_{1})^{\circledast}y_{13}%
+y_{12}^{\circledast}y_{23}-y_{13}^{\circledast}(x-y_{3})]\nonumber\\
&  \times\lbrack\lambda_{3}^{\circledast}\lambda_{2}+y_{13}^{\circledast
}y_{12}-y_{23}^{\circledast}(x-y_{2})-(x-y_{3})^{\circledast}y_{23}%
]\nonumber\\
&  \times\lbrack\lambda_{2}^{\circledast}\lambda_{1}-y_{12}^{\circledast
}(x-y_{1})-(x-y_{2})^{\circledast}y_{12}+y_{23}^{\circledast}y_{13}%
]\nonumber\\
&  -[|\lambda_{1}|_{c}^{2}+|x-y_{1}|_{c}^{2}+|y_{12}|_{c}^{2}+|y_{13}|_{c}%
^{2}]\nonumber\\
&  \times\lbrack\lambda_{2}^{\circledast}\lambda_{3}+y_{12}^{\circledast
}y_{13}-(x-y_{2})^{\circledast}y_{23}-y_{23}^{\circledast}(x-y_{3}%
)]\nonumber\\
&  \times\lbrack\lambda_{3}^{\circledast}\lambda_{1}-y_{13}^{\circledast
}(x-y_{1})+y_{23}^{\circledast}y_{12}-(x-y_{3})^{\circledast}y_{13}%
]\nonumber\\
&  -[\lambda_{1}^{\circledast}\lambda_{2}-(x-y_{1})^{\circledast}y_{12}%
-y_{12}^{\circledast}(x-y_{2})+y_{13}^{\circledast}y_{23}]\nonumber\\
&  \times\lbrack\lambda_{2}^{\circledast}\lambda_{1}-y_{12}^{\circledast
}(x-y_{1})-(x-y_{2})^{\circledast}y_{12}+y_{23}^{\circledast}y_{13}%
]\nonumber\\
&  \times\lbrack|\lambda_{3}|_{c}^{2}+|x-y_{3}|_{c}^{2}+|y_{13}|_{c}%
^{2}+|y_{23}|_{c}^{2}]\nonumber\\
&  -[\lambda_{1}^{\circledast}\lambda_{3}-(x-y_{1})^{\circledast}y_{13}%
+y_{12}^{\circledast}y_{23}-y_{13}^{\circledast}(x-y_{3})]\nonumber\\
&  \times\lbrack|\lambda_{2}|_{c}^{2}+|x-y_{2}|_{c}^{2}+|y_{12}|_{c}%
^{2}+|y_{23}|_{c}^{2}]\nonumber\\
&  \times\lbrack\lambda_{3}^{\circledast}\lambda_{1}-y_{13}^{\circledast
}(x-y_{1})+y_{23}^{\circledast}y_{12}-(x-y_{3})^{\circledast}y_{13}%
]\nonumber\\
&  =0. \label{jump3}%
\end{align}
The result of the determinant in Eq.(\ref{jump3}) corresponds to a polynomial
of degree $2k=6$ in the variable $x$. The number of parameters in
Eq.(\ref{jump3}) can be reduced to $42$ by making the choices of parameters in
the paragraph after Eq.(\ref{3-ins}) and using Eq.(\ref{1}) to Eq.(\ref{5}).
For the special case of $y_{ij}=0$ $(i\neq j)$, $\lambda_{i}$ reduces to a
complex number $\lambda_{i}=\lambda_{i}e_{0}$ (with $\lambda_{i}$ a
\textit{complex }number) and Eq.(\ref{jump3}) reduces to Eq.(\ref{k-jump}).

For $x\in J$ in the jumping lines of $SL(2,C)$ $k$-instantons, the
non-singular property in the quadratic condition of ADHM construction in
Eq.(\ref{ff}) is violated. Moreover, the projection operator $P$ in
Eq.(\ref{P}) and the self-dual field strength $F_{\mu\nu}$ in Eq.(\ref{F})
diverge there. For some cases of these singularities, the holomorphic vector
bundle description of the original ADHM construction for $SU(2)$ instantons
fails and one is led to include sheaf in the construction of the non-compact
$SL(2,C)$ $k$-instantons. It can be shown mathematically that \cite{math2}
there are solutions of complex $SL(2,C)$ ADHM equations \cite{Donald} which
admit jumping lines. In this paper we have given explicit concrete examples to
describe jumping lines of $SL(2,C)$ $k$-instantons from physicist point of
view. More examples of $SL(2,C)$ instanton connections corresponding to
sheaves with jumping lines instead of vector bundles on $CP^{3}$ are under
investigation \cite{LLT}.

\section{Conclusion}

In this paper, we have extended the ADHM construction of $Sp(1)$ self-dual
Yang-Mills (SDYM) instantons to the case of $SL(2,C)$ instanton solutions. In
constrast to the quaternion calculation heavily used in the compact groups
$Sp(N)$, $SU(N)$ and $O(N)$ constructions in the literature \cite{CSW}, we
discover that instead the use of biquaternion with the biconjugation operation
\cite{Ham} is very powerful in the construction of the non-compact $SL(2,C)$
instanton solutions. The new $SL(2,C)$ instanton solutions constructed by this
$SL(2,C)$ ADHM method contain those $SL(2,C)$ $(M,N)$ instanton solutions
\cite{Lee} obtained previously as a subset. We found that the number of moduli
for $SL(2,C)$ $k$-instantons is twice of that of $Sp(1)$, $16k-6$.

In addition, we investigate the structure of singularities or jumping lines
\cite{math2} of the complete $SL(2,C)$ $1$-instanton solution with $10$ moduli
parameters. The singularities are intersections of zeros of $P_{2}(x)$ and
$P_{1}(x)$ polynomials of $4$ variables with degrees $2$ and $1$ respectively.
For singularities of subsets of $k$-instanton field cofigurations, one
encounters intersections of zeros of $P_{2k}(x)$ and $P_{2k-1}(x)$ polynomials
with degrees $2k$ and $2k-1$ respectively. We found that not all singularities
can be gauged away as in the case of $SU(2)$ $1$-instanton field
confuguration. The singularities for $SL(2,C)$ $1$-instanton field
configuration is much more complicated than the removable singularity of
$SU(2)$ $1$-instanton field configuration. Moreover, the values of all $10$
parameters are closely related to the structure of the singularities.

The jumping lines of the complete $SL(2,C)$ $k=2,3$ instantons with $26,42$
moduli parameters are also calculated in this paper. Mathematically, the
existence of singular structures of the non-compact $SL(2,C)$ $k$-instanton
field configurations discovered in this paper is consistent with the inclusion
of "sheaves" by Frenkel-Jardim \cite{math2}, rather than just the restricted
notion of "vector bundles", in the one to one correspondence between ASDYM and
certain algebraic geometric objects. In this paper we have given explicit
concrete examples to describe "jumping lines" of $SL(2,C)$ $k$-instantons from
physicist point of view.

The existence of non-removable singular structures of $SL(2,C)$ instanton
field configurations may help to clearify the long standing issue of global
singularity problems associated with Backlund transformations \cite{PBT,AW} of
$SU(2)$ SDYM instantons. Further investigation of the structure of
singularities for general $SL(2,C)$ $k$-instanton field configurations maybe
important for the understanding of the geometrical structures of non-compact
SDYM theory.

\section{Acknowledgments}

J.C. Lee would like to thank R. Sasaki and C.C. Hsieh for discussions. The
work of J.C. Lee is supported in part by the Ministry of Science and
Technology, National Center for Theoretical Sciences and S.T. Yau center of
NCTU, Taiwan. I-H. Tsai would like to thank S.T.Yau center of NCTU for the
invitation and the stimulating atmosphere where part of his work was done.

\end{document}